\documentclass[11pt]{article}
\pdfoutput=1
\usepackage{jcapmod}
\usepackage{booktabs}
\usepackage[english]{babel}
\usepackage{amsmath,amssymb,amsbsy,amstext, amsthm, simplewick}
\usepackage{graphicx}
\usepackage{amsfonts}
\usepackage{amssymb}
\usepackage{upgreek}
 \usepackage{exscale,relsize}
 \usepackage[makeroom]{cancel}
\usepackage{soul}
\usepackage{slashed}

\usepackage{tensor}

\def\PB{ \mathcal{P}_B}
\def\PE{ \mathcal{P}_E}

\def\beq{\begin{equation}}
\def\eeq{\end{equation}}
\def\bea{\begin{eqnarray}}
\def\eea{\end{eqnarray}}

\numberwithin{equation}{section}

\begin{document}

\providecommand{\abs}[1]{\lvert#1\rvert}
\providecommand{\bd}[1]{\boldsymbol{#1}}

\begin{titlepage}

\setcounter{page}{1} \baselineskip=15.5pt \thispagestyle{empty}

\bigskip\

\vspace{1cm}

\begin{center}

{\fontsize{20}{28}\selectfont  \sffamily \bfseries Constraints on  Primordial }
\vskip 12pt
{\fontsize{20}{28}\selectfont  \sffamily \bfseries Magnetic Fields from Inflation}

\end{center}

\vspace{0.2cm}

\begin{center}
{\fontsize{14}{30}\selectfont   Daniel Green$^{\star, \ddagger}$
and
Takeshi Kobayashi$^{\star, \ast, \dagger}$ }
\end{center}

\begin{center}
\textsl{$^{\star}$ Canadian Institute for Theoretical Astrophysics,
 University of Toronto, \\ 60 St. George Street, Toronto, Ontario M5S
 3H8, Canada}\\

\vskip 6pt
\textsl{$^{\ddagger}$ Canadian Institute for Advanced Research, Toronto,
 Ontario M5G 1Z8, Canada}\\

\vskip 6pt
\textsl{$^{\ast}$ Perimeter Institute for Theoretical Physics, \\ 
 31 Caroline Street North, Waterloo, Ontario N2L 2Y5, Canada}\\

\vskip 6pt
\textsl{$^{\dagger}$ SISSA, Via Bonomea 265, 34136 Trieste, Italy}\\


\end{center} 



\vspace{1cm}

\vspace{1.2cm}
\hrule \vspace{0.3cm}
\noindent {\sffamily \bfseries Abstract} \\[0.1cm]
We present generic bounds on magnetic fields produced from cosmic
 inflation. By investigating field bounds on the vector potential, we
 constrain both the quantum mechanical production of magnetic
 fields and their classical growth in a model independent way. For classical growth, we show
 that only if the reheating temperature is as low as
 $T_{\mathrm{reh}} \lesssim 10^2\, \mathrm{MeV}$ can magnetic fields
 of $10^{-15}\, \mathrm{G}$ be produced on Mpc scales in the present
 universe. For purely quantum mechanical scenarios, even stronger constraints
 are derived. Our bounds on classical and quantum mechanical scenarios
 apply to generic theories of inflationary magnetogenesis with a
 two-derivative time kinetic term for the vector potential.  In both cases, the magnetic field strength is 
 limited by the gravitational back-reaction of the electric fields that are produced simultaneously. As an example of
 quantum mechanical scenarios, we construct vector field theories whose
 time diffeomorphisms are spontaneously broken, and explore magnetic
 field generation in theories with a variable speed of
 light. Transitions of quantum vector field fluctuations into classical
 fluctuations are also analyzed in the examples.  
\vskip 10pt
\hrule

\vspace{0.6cm}
 \end{titlepage}

\tableofcontents

\newpage

\section{Introduction}
\label{sec:intro}

Magnetic fields exist in our universe on various length scales, however 
their origin remains a mystery. One possibility that has been vigorously
pursued is that the magnetic fields were seeded in the early universe,
in particular, during the inflationary epoch.
The purpose of this paper is to examine the validity of inflationary
magnetogenesis in a model independent way, 
by analyzing field bounds for the vector potential.

There are two ways to have large magnetic fields by the end of
cosmic inflation:
\begin{itemize}
 \item {\it Classical growth.} Initially tiny magnetic fields (such as
       those originating from 
       quantum fluctuations of the vector potential) undergo classical
       growth during inflation.
 \item {\it Quantum mechanical production.} Large vector field
       fluctuations are produced at the quantum 
       level, setting large magnetic fields as an `initial condition'
       for the subsequent classical evolution.        
\end{itemize}
While many authors have investigated the former scenario since the seminal
works~\cite{Turner:1987bw,Ratra:1991bn}, the
latter is much less studied. 
Here we note that quantum mechanical scenarios can also be studied in 
a computable framework.
For instance, if the vector field has a non-trivial dispersion relation during
inflation then the fields can, in principle, obtain large quantum fluctuations
leading to large magnetic fields. 
This can be achieved in vector field theories with higher
derivatives, or with spontaneously broken Lorentz invariance.
Alternatively, an oscillating background can also induce quantum mechanical
production of the vector through parametric resonance.

For both classical and quantum mechanical scenarios,
in this paper we derive upper bounds 
on the magnetic fields that can be produced during inflation,
independently of the details of the model.
To achieve this, we study field range bounds on the vector
potential.\footnote{In this sense, our approach 
can be related to the Lyth bound~\cite{Lyth:1996im} in inflation. 
However, unlike the Lyth ``bound" which connects observably large gravitational
waves to super-Planckian field excursions of the inflaton field,  
our magnetic field bounds will arise from the requirement that
the vector field's energy density should not significantly backreact on
the inflationary background.}
For classical scenarios, magnetic field generation is supported by the
time-evolution of the vector potential. 
By analyzing the induced kinetic energy of the vector field (which
is proportional to the electric field strength squared)
and its backreaction to the background cosmology, 
we constrain the vector field evolution during inflation.
The calculations will be carried out independently of the specific
dynamics of the vector field.

Quantum mechanical scenarios, on the other hand, do not necessarily rely on 
time-evolving vector fields, but instead invoke large quantum
fluctuations.
Nevertheless, magnetogenesis in these scenarios are also found to be
severely constrained, as the energy that can be extracted from the
inflationary background is limited.
By computing the background energy that is available for the quantum
fluctuations,
we constrain magnetic fields that are set as `initial
conditions' for the subsequent classical evolution.
Upon discussing quantum mechanical scenarios, we will also study
how the quantum vector fluctuations transform into classical
fluctuations in some example models.

We will occasionally compare our magnetic field
bounds with the lower limit on intergalactic magnetic fields reported
by recent gamma ray observations, which is typically of 
$10^{-15}\, \mathrm{G}$ assuming a correlation length of~$\lambda_B \geq 1\,
\mathrm{Mpc}$~\cite{Tavecchio:2010mk,Neronov:1900zz,Ando:2010rb,Taylor:2011bn,Takahashi:2013lba,Chen:2014rsa}.
(If $\lambda_B$ is much smaller than a Mpc, the lower limit of the intergalactic
magnetic field strength further improves as~$\propto \lambda_B^{-1/2}$.)
Our results will show that, within our assumptions, only for a very narrow window
of the inflation and reheating scales is it possible to produce 
magnetic field strengths of $10^{-15}\, \mathrm{G}$ on Mpc scales or larger.

In our calculations for constraining inflationary
magnetogenesis, we will assume that the vector field theory has a
two-derivative time kinetic term; 
within this assumption, our bounds apply to generic vector field
theories, including models with interactions with other physical degrees of
freedom, broken time diffeomorphisms, 
higher-order spatial derivatives, etc.
As an example of quantum mechanical magnetogenesis, 
we also perform an analysis of vector field theories in backgrounds that break the time diffeomorphisms. 
The theory is studied using similar techniques
to~\cite{Creminelli:2006xe,Cheung:2007st} for formulating the effective 
field theory of inflation;
we start by constructing the vector field action in unitary gauge. 
When the gauge invariant action (i.e. the action whose time
diffeomorphisms are restored) is needed for computing quantities such as
the energy-momentum tensor, we will use the St\"uckelberg trick to
reintroduce the Nambu-Goldstone boson of the broken time
diffeomorphisms. 
We then discuss magnetic field generation in cases where the broken Lorentz
symmetry gives rise to a variable speed of light.

The paper is organized as follows: We first define the electromagnetic
fields and set our notation in Section~\ref{sec:EMfields}. 
Then we derive magnetic field bounds on classical and quantum mechanical
scenarios in 
Section~\ref{sec:time-evolution} and Section~\ref{sec:lqf},
respectively. 
In Section~\ref{sec:conc} we summarize our findings and 
discuss the implications for cosmological magnetogenesis. 
The appendices are devoted to studies of explicit models.
In Appendix~\ref{app:IFF}, we study $I^2 FF$ models as a typical example
of the classical scenarios discussed in Section~\ref{sec:time-evolution}.
In Appendix~\ref{app:BTD}, we construct and study a vector field theory
with a general light speed.
Here we also analyze how the quantum vector fluctuations transform into
classical fluctuations. 
The model investigated in this appendix will serve as an example of quantum
mechanical scenarios discussed in Section~\ref{subsec:smooth}.

\section{Electromagnetic Fields}
\label{sec:EMfields}

The theory of electromagnetism could have been
quite different in the early universe, 
e.g. with higher-derivative terms for the vector potential, 
interactions with beyond the Standard Model particles such as the
inflaton. 
However by the present times, the
theory should recover the standard Maxwell theory:
\begin{equation}
 S = \int d^4 x \sqrt{-g} \left\{
-\frac{1}{4} F_{\mu \nu} F^{\mu \nu} + 
(\text{couplings to charged fields})
\right\},
\label{standMaxwell}
\end{equation}
where $F_{\mu \nu} = \partial_{\mu} A_{\nu} - \partial_\nu A_\mu$.
We use Greek letters for the spacetime indices $\mu, \nu = 0, 1, 2, 3$,
and Latin letters for spatial indices $i, j = 1,2,3$. 
Then we can define the electric and magnetic fields such that by the
present times, they become the fields measured 
by a comoving observer with
$4$-velocity~$u^{\mu}$ ($u^i = 0$, $u_\mu u^{\mu} = -1$),
\begin{equation}
E_{\mu} = u^\nu F_{\mu \nu} ,
\qquad
B_{\mu} = \frac{1}{2} \varepsilon_{\mu \nu \sigma} F^{\nu \sigma},
\label{EBformal}
\end{equation}
where 
$\varepsilon_{\mu \nu \sigma} = \eta_{\mu \nu \sigma \rho} u^\rho$,
and $\eta_{\mu \nu \sigma \rho}$ is a totally antisymmetric tensor
with $\eta_{0123} = -\sqrt{-g}$.  Throughout this paper we consider a flat FRW universe,
\begin{equation}
  ds^2 = a(\tau)^2 \left( -d\tau^2 + d\bd{x}^2 \right). 
\label{FRW}
\end{equation}
Metric fluctuations will not play an important role in our discussion and have been set to zero.  

We assume the electromagnetic theory preserves the gauge invariance
$A_\mu \to A_\mu - \partial_\mu \theta$ 
at all times, 
thus it is possible to always choose a gauge in which the time component
of the vector potential vanishes,
\begin{equation}
 A_0 = 0.
\end{equation}
Then the electromagnetic field amplitudes are written as
\begin{equation}
 E^\mu E_\mu = \frac{1}{a^4} A_i' A_i',
\qquad
 B^\mu B_\mu = \frac{1}{a^4}
\left( 
\partial_i A_j \partial_i A_j
- 
\partial_i A_j \partial_j A_i
\right),
\label{E2B2}
\end{equation}
where a prime denotes a derivative in terms of the comoving
time~$\tau$, 
and the sum over repeated spatial indices is implied
irrespective of their positions. 

\vspace{\baselineskip}

In the case of the standard Maxwell theory~(\ref{standMaxwell}),
when ignoring couplings to other fields and taking the Coulomb gauge $A_0 =
\partial_i A_i = 0$, the action reduces to (after dropping surface terms),
\begin{equation}
 S = \frac{1}{2}  \int d \tau d^3 x 
\left( 
A_i^{\mathrm{T}\prime}  A_i^{\mathrm{T}\prime} - \partial_i A_j^\mathrm{T} \partial_j A_i^\mathrm{T}
\right).
\label{Max2.6}
\end{equation}
Here the ``T'' superscript denotes the vector potential being transverse. 
The equations of motion, i.e. Maxwell's equations, read
\begin{equation}
A_i^{\mathrm{T}\prime \prime} - \partial_j \partial_j A_i^\mathrm{T} = 0,
\end{equation}
yielding plane wave solutions. 
Thus one sees that both the electric and magnetic field strengths
squared~(\ref{E2B2}) basically decay as $\propto a^{-4}$ 
in an expanding universe.  Furthermore, Maxwell theory is conformal 
and therefore the evolution of $A_i$ is equivalent to in flat space, as we can see from (\ref{Max2.6}).  

In order for large magnetic fields to arise from the early universe, 
the Maxwell theory has to be modified.  Conformal invariance must necessarily be broken to seed the classical primordial fluctuations from the (quantum) vacuum.  
Furthermore, we must alter fields'
time evolution and/or amplitude of primordial fluctuations in order to leave appreciable magnetic fields at late times.
We will discuss each of these possibilities in the following sections and show that, in either case, 
magnetic field production during the inflationary era is highly restricted.

\section{Constraints on Classical Growth}
\label{sec:time-evolution}

Let us study magnetogenesis scenarios where the initially
tiny magnetic fields, such as those originating from quantum fluctuations
of the vector potential, experience classical growth during the
inflationary epoch.
Examples of such scenarios are~\cite{Turner:1987bw,Ratra:1991bn}, which
considered breaking the conformal invariance of the Maxwell theory
during inflation;
after the wave modes exit the effective ``horizon'' induced by the
conformal symmetry breaking, the modes become classical and 
the cosmological background can enhance the 
super-horizon magnetic fields.
However, detailed studies of individual models have revealed that, for the
simplest conformal symmetry breaking scenarios, electric fields are also
enhanced along with the 
magnetic fields and tend to become so large as to spoil inflation and/or
magnetogenesis~\cite{Bamba:2003av,Demozzi:2009fu,Kanno:2009ei,Fujita:2012rb,Kobayashi:2014sga}.
In this section, we discuss the electric field production independently
of the details of the model, and show that the electric backreaction
problem is inherent to the classical growth of magnetic fields during inflation.
The approach we take is close to the 
one presented in~\cite{Fujita:2012rb} for studying $I^2 FF$ models,
however, we obtain a much
stronger result by optimizing the combination of constraints in the
theory.  
For the differences with the calculations in~\cite{Fujita:2012rb}, 
we refer the reader to Appendix~\ref{app:IFF}.

In order to discuss electromagnetic fields with certain 
correlation lengths, it is convenient to go to Fourier space,
\begin{equation}
 A_i (\tau, \bd{x}) = \frac{1}{ (2 \pi)^3}\int d^3 k \, 
e^{i \bd{k \cdot x}} \tilde{A}_{i,k} (\tau) a_{\bd{k}} ,
\label{fourierA_i}
\end{equation}
where $a_{\bd{k}}$ is a stochastic variable that satisfies $\langle
a_{\bd{k}} a^*_{\bd{k}'} \rangle = (2 \pi)^3 \delta(\bd{k}-\bd{k}')$.
(In the quantum mechanical case, 
as in Section~\ref{sec:lqf}, the stochastic variables $a_{\bd{k}}$ and $a_{\bd{k}}^*$
are promoted to annihilation and creation operators.)  This
definition\footnote{Given a measurement of a electric or magnetic field,
one would use a discrete Fourier transform to describe the field rather than the continuous one that
appears here.  For the discrete case, we should use $\langle a_{i}
a^*_{j} \rangle = \delta_{ij}$ where $i$ labels the discrete momentum
vectors.  This is important only for establishing the correct units for
$A_{i,k}$ in each case. We also note that, strictly speaking, we should
introduce the stochastic variables for each independent degrees of freedom
of the vector field; however the expression~(\ref{fourierA_i}) is
sufficient for the purpose of obtaining order-of-magnitude estimates.} is meant to reflect the idea that the initial amplitude for $\tilde{A}_{i,k}$ arises from a random process.
We also remark that since we are interested in an FRW background,
the amplitude $\tilde{A}_{i,k}$ is considered to depend only on the
magnitude $k = \abs{\bd{k}}$ of the comoving wave number.

We evaluate the electric and magnetic field strengths~(\ref{E2B2})
associated with~$k$ in terms of the power spectra defined as
\bea
\langle B_\mu (\tau, \boldsymbol{x}) B^\mu (\tau, \boldsymbol{y}) \rangle &=&
 \int \frac{d^3 k}{4 \pi k^3}  e^{i\boldsymbol{k\cdot}  (\boldsymbol{x
- y})} \mathcal{P}_B (\tau, k) \ , \\
\langle E_\mu (\tau, \boldsymbol{x}) E^\mu (\tau, \boldsymbol{y}) \rangle &=&
 \int \frac{d^3 k}{4 \pi k^3}  e^{i\boldsymbol{k\cdot}  (\boldsymbol{x
- y})} \mathcal{P}_E (\tau, k) \ .
\eea
Here the factor of $k^{-3}$ is a convention such that $\PB$ and $\PE$
have the same units as $B^2$ and $E^2$.
Noting that $A_i$ is real, 
the power spectra can be obtained as 
\begin{gather}
 \mathcal{P}_B (\tau, k) = \frac{k^3}{2 \pi^2 a(\tau)^4}
  \left\{
   k^2 \tilde{A}_{i,k} (\tau)\tilde{A}_{i,k}^* (\tau)
-   \abs{ k_i \, \tilde{A}_{i,k} (\tau)}^2
	\right\}
  \leq \frac{k^5}{2 \pi^2 a(\tau)^4}
 \tilde{A}_{i,k} (\tau)\tilde{A}_{i,k}^* (\tau) ,
 \label{eq:PB}
\\
 \mathcal{P}_E (\tau, k) = \frac{k^3}{2 \pi^2 a(\tau)^4} 
 \tilde{A}'_{i,k} (\tau) \tilde{A}_{i,k}'^* (\tau)  \ .
\label{eq:PE}
\end{gather}
Since we are interested in order-of-magnitude estimates,
in the following we ignore the component subscript~$i$.

Large magnetic fields can emerge if $\abs{\tilde{A}_k}$ grows in time
during inflation, hence let us evaluate the 
time-evolution of the vector potential as
\begin{equation}
 \abs{\tilde{A}_{k}(\tau_f)} -  \abs{\tilde{A}_k(\tau_i)}
  = \int^{\tau_f}_{\tau_i} d\tau \, \frac{d \abs{\tilde{A}_k}}{d\tau }
  \leq
  \int^{\tau_f}_{\tau_i} d\tau \, \left| \frac{d \tilde{A}_k}{d\tau } \right|.
  \label{eq9}
\end{equation}
Here, $\tau_f$ is the time during or at the end of inflation
when the time-evolution of~$\abs{\tilde{A}_k}$ ceases,
and $\tau_i$ is some arbitrary time before~$\tau_f$.
Upon moving to the far right hand side, we used the
inequality $d \abs{f(x)} / dx \leq \abs{ df(x) / dx }$ which holds for
any\footnote{This inequality is ill-defined when $f(x)$ crosses zero.  Nevertheless, the integrated inequality still holds provided $f(x)$ is differentiable.} complex function~$f(x)$ with a real variable~$x$.

The vector field action during inflation may contain higher-derivatives
and/or interaction terms.
However, let us suppose that the dominant time kinetic term of the
vector field during $\tau_i \leq \tau \leq \tau_f$ is a two-derivative term,
\begin{equation}
 S = \int d\tau d^3 x \left(
\frac{I^2}{2}  A_i' A_i' + \cdots
\right).
\label{2der-kin}
\end{equation}
Here, $I^2$ is a dimensionless parameter which is defined to be $I =1$ in the present epoch.  This definition implies that
 $A_i$ becomes canonically normalized and the
theory connects to the standard Maxwell theory~(\ref{Max2.6}) at late times.
For example, in the model proposed in~\cite{Ratra:1991bn}, $I^2$ is a
function of the inflaton field and thus varies in time as the inflaton
field rolls along its potential.

The parameter~$I^2$ can in principle take arbitrary values during
inflation, and smaller values are preferable for avoiding backreaction
issues, as we will soon see. 
However it
should be noted that $I^2 \ll 1$ would push the theory
into the strong coupling regime where perturbative calculations break
down~\cite{Demozzi:2009fu,Gasperini:1995dh}: 
Considering an interaction with a charged fermion such as $e
\bar{\psi} \gamma^\mu \psi A_\mu$, then after canonically normalizing
the vector field as $A^c_\mu = I A_\mu$, the interaction term becomes 
$\frac{e}{I} \bar{\psi} \gamma^\mu \psi A^c_\mu $ and thus one sees that
the effective coupling is $e/I$.
We also remark that, if $I^2 \ll 1 $ 
throughout inflation, then one should further analyze the vector field dynamics
after inflation as $I^2$ comes back to unity.  

Alternatively, we are free to work with the canonically normalized
field, $A^c_\mu$, at all times. While this is equivalent to setting $I
\equiv 1$ for the standard kinetic term, for a general time dependent
$I(\tau)$ this will introduce additional terms in the action
proportional to $I' / I$.  If these terms are small relative to the two
derivative kinetic term, then these additional terms do not violate any
of our technical assumptions below that the results will hold even if $I \ll
1$ in the past. As a result, $I \ll 1$ is only a meaningful possibility
when $|I'/I|$ is sufficiently large that these terms are comparable in size to the standard kinetic term.  Of course, even in this case, one has to worry about the strong coupling constraint as well.  (See also Section~\ref{sec:conc} for related discussions.)

In the following, to avoid the aforementioned issues, we require
\begin{equation}
 I^2 \gtrsim 1
\label{coupling-req}
\end{equation}
to hold during the times $\tau_i \leq \tau \leq \tau_f$.  For a classical vector field, the two-derivative kinetic term
contributes to the field's energy density as
\begin{equation}
 \rho_{\mathrm{kin}} \sim
  \left\langle
\frac{I^2}{a^4}A_i' A_i'
\right\rangle
=
\int \frac{dk}{k} 
\frac{k^3}{2 \pi^2} \frac{I^2}{a^4} |\tilde{A}'_{k}|^2  
\label{eq3.6}
\end{equation}
which is roughly the electric field strength squared multiplied by the
coefficient~$I^2$, cf.~(\ref{eq:PE}). 
(For a quantum field, this estimate of the energy density is 
not necessarily meaningful as the contributions from high $k$ modes should be
renormalized and absorbed into the definition of the cosmological constant.)
Supposing that the kinetic contribution~(\ref{eq3.6}) to the vector field's energy
density is not cancelled by other vector terms in the action, then
$\rho_{\mathrm{kin}}$ should be much smaller than the total energy density of the
inflationary universe
to avoid the vector fields from spoiling inflation and/or
magnetogenesis.
Thus, for purely classical modes we require
\begin{equation}
 \int \frac{dk}{k}  \frac{k^3}{2 \pi^2} \frac{I^2}{a^4} |\tilde{A}'_{k}|^2 \ll 3 M_p^2 H_{\mathrm{inf}}^2,
\label{energy3.6}
\end{equation}
where $H_{\mathrm{inf}}$ is the inflationary Hubble rate.

Now, let us assume that the vector field fluctuations on wave modes of
interest become classical by the time~$\tau_i$.\footnote{In conformal
symmetry breaking models such 
as~\cite{Turner:1987bw,Ratra:1991bn}, the vector fluctuations become
classical after the modes exit the effective horizon induced by the
symmetry breaking. 
However we also note that modes can become classical much earlier, 
if, for instance, an oscillatory background gives rise to resonant
production of the vector field while the modes are deep inside the
horizon. 
The initial time~$\tau_i$ is considered to be taken after the mode is effectively classical.
See also the appendices, in particular Appendix~\ref{subsec:cl-qu},
where we discuss how modes become classical in more detail.}
By `becoming classical,' we mean that
the amplitude of the commutator of the vector field and its conjugate momentum 
$[\tilde{A}_k, \tilde{\Pi}_h]^2$
becomes negligibly tiny compared to
$\langle \tilde{A}_k \tilde{A}_h \rangle
\langle \tilde{\Pi}_k \tilde{\Pi}_h \rangle $ (see e.g.~\cite{Maldacena:2015bha} for a recent discussion).
Then from the above discussions, the contribution to the kinetic energy
from each classical mode is bounded by the background density as
\begin{equation}
 \frac{k^3}{2 \pi^2} \frac{I^2}{a^4} |\tilde{A}'_{k}|^2
\ll 3 M_p^2 H_{\mathrm{inf}}^2 \ ,
\label{energy-req}
\end{equation}
where we have assumed that there are modes populating a range of modes,
$\Delta k \simeq k$, such that we can approximate the integral as $\int
\frac{dk}{k} \to 1$.  This condition constrains the time derivative of
the vector potential.\footnote{The condition~(\ref{energy-req}) could be violated,
if the spectrum of $k^3 \abs{\tilde{A}_k'}$ has sharp features localized
to a very narrow range of $k$ such that $\Delta k \ll k$.
In order to achieve such a narrow window, the action would require
rapidly time varying coefficients which will typically lead to particle
production.  It is likely that such scenarios would be characterized by
the discussion in Section~\ref{subsubsec:res},
but in principle there could be room to violate (\ref{energy-req}) and
produce large magnetic fields over very narrow $k$~ranges.
In such cases, one should instead discuss
the magnetic field spectrum integrated over wave numbers of interest
in order to relate to measurements of magnetic fields; 
see also Footnote~\ref{footinapp}.\label{foot4}}
Therefore the combination of (\ref{eq9}),
(\ref{coupling-req}), and (\ref{energy-req}) yields
\begin{equation}
 \abs{\tilde{A}_k(\tau_f)} -  \abs{\tilde{A}_k(\tau_i)}
  \lesssim
  \int^{\tau_f}_{\tau_i} d\tau \,  \frac{\sqrt{6} \pi a^2 M_p
  H_{\mathrm{inf}} }{ k^{3/2}}
  = \frac{\sqrt{6} \pi M_p \left( a_f - a_i \right)}{ k^{3/2}} .
\end{equation}
Here, upon obtaining the far right hand side we have used the relationship
$d \tau = da / (a^2 H_{\mathrm{inf}})$ for the conformal time in a de
Sitter universe, and $a_f = a(\tau_f)$, $a_i = a(\tau_i)$.
In this section we are interested in cases where the vector potential
grows after becoming classical,
therefore, taking the time~$\tau_i$ to be sufficiently before
the end of magnetogenesis such that $a_f \gg a_i$, 
we can suppose
\begin{equation}
 \abs{ \tilde{A}_k (\tau_f) } \gg  \abs{ \tilde{A}_k (\tau_i) },
    \label{growing_assump}
\end{equation}
and obtain an upper bound on the absolute value of the vector potential,
\begin{equation}
\abs{ \tilde{A}_k (\tau_f) } 
 \lesssim
 \frac{ \sqrt{6} \pi M_p \, a_f }{k^{3/2}}\ .
\label{Akbound}
\end{equation}
Since $\abs{ \tilde{A}_k } $ becomes time independent after~$\tau_f$ and
the magnetic field~(\ref{eq:PB}) redshifts as $\PB \propto a^{-4} $,
we obtain an upper bound on the magnetic field strength in the
present universe,
\begin{equation}
\PB(\tau_0, k) = \PB(\tau_f, k)   \left( \frac{a_f}{a_0} \right)^4
 \lesssim 
  3 M_p^2 \left( \frac{k}{a_0} \right)^2 \left(
	 \frac{a_f}{a_0} \right)^2
\leq  3  M_p^2 \left( \frac{k}{a_0} \right)^2 
  \left(\frac{ a_{\mathrm{end}}}{a_0} \right)^2.
  \label{eq15}
\end{equation}
The subscript ``0'' denotes values in the present universe, and $a_{\mathrm{end}} $ ($\geq a_f$) is the scale factor at the end of inflation. 

Let us suppose that the universe is effectively matter-dominated after
inflation until reheating, i.e.,
\begin{equation}
 \left( \frac{H_{\mathrm{reh}}}{H_{\mathrm{inf}}} \right)^2 =
 \left( \frac{a_{\mathrm{end}}}{a_{\mathrm{reh}}} \right)^3,
\label{Hrehinf}
\end{equation}
where values at reheating are represented by the subscript ``reh.''
Then, considering entropy conservation ($s \propto a^{-3}$) after reheating,
we have\footnote{To compute the entropy density
at reheating 
\begin{equation}
s_{\mathrm{reh}} = \frac{2 \pi^2}{45}g_{s*}(T_{\mathrm{reh}})
\left( \frac{90}{\pi^2}\frac{ M_p^2 H_{\mathrm{reh}}^2}{g_*
 (T_{\mathrm{reh}})} \right)^{3/4} ,
\end{equation}
we have chosen the relativistic degrees of freedom to take the 
maximum value allowed in the MSSM, $g_* = g_{s*} = 228.75$.
However we note that this choice affects~(\ref{HrehMp}) only by an
order unity factor; e.g., $g_* = g_{s*} = 10.75$ gives
a factor~$7$ instead of~$6$ in the right hand side.}
\begin{equation}
 \frac{a_{\mathrm{reh}}}{a_0} \approx 6 \times 10^{-32} 
\left( \frac{M_p}{H_{\mathrm{reh}}} \right)^{1/2},
\label{HrehMp}
\end{equation}
which together with (\ref{Hrehinf}) gives the expansion after inflation,
\begin{equation}
 \frac{a_{\mathrm{end}}}{a_0}
\approx 6 \times  10^{-32}
\left(\frac{H_{\mathrm{reh}}}{H_{\mathrm{inf}}}\right)^{1/6}
 \left(\frac{M_p}{H_{\mathrm{inf}}}\right)^{1/2}.
\label{rehinf63}
\end{equation}
Further using 
$1\,  G \approx 2.0 \times 10^{-20} \, \mathrm{GeV}^2 $ (we use the Heaviside-Lorentz
units) and $1\, \mathrm{Mpc} \approx 1.6 \times 10^{38}\, \mathrm{GeV}^{-1}  $,
then the magnetic field bound (\ref{eq15}) is rewritten as
\begin{equation}
\PB(\tau_0, k)  \lesssim 
(10^{-15}\, \mathrm{G})^2  
 \left(\frac{k}{a_0}\, \mathrm{Mpc}\right)^2
 \left( \frac{H_{\mathrm{reh}}}{H_{\mathrm{inf}}} \right)^{1/3}
 \frac{10^{-14}\, \mathrm{GeV}}{H_{\mathrm{inf}}}.
 \label{B0bound1}
\end{equation}
Thus we find that in order to create magnetic fields of $10^{-15}\,
\mathrm{G}$ on Mpc scales or larger, the inflation scale has to
be as low as $H_{\mathrm{inf}} \lesssim 10^{-14} \, \mathrm{GeV}$;
otherwise the vector field's electric backreaction to the inflationary universe
would be significant, and/or the theory would have strong couplings. 
The bound becomes even stronger when there is a hierarchy between the
inflation and reheating scales. 

Upon obtaining (\ref{B0bound1}), we have 
only required the kinetic energy of the vector field to be smaller than
the total energy density of the inflationary universe~$ \rho_{\mathrm{inf}}$,
cf.~(\ref{energy3.6}). 
Here we should remark that this condition is not sufficient when
taking into account the cosmological perturbations, as the vector field
itself sources curvature perturbations of
roughly~\cite{Barnaby:2012tk,Suyama:2012wh,Giovannini:2013rme,Fujita:2014sna,Ferreira:2014hma}
\begin{equation}
 \zeta_{A} \sim \frac{1}{\epsilon } 
\frac{\rho_{A}}{\rho_{\mathrm{inf}}},
\label{zeta_A}
\end{equation}
where $\rho_{\mathrm{A}}$ is the vector field's energy density and 
$\epsilon = - H'/(a H^2)$ is the rate of change of the Hubble parameter
during inflation. 
Since (\ref{zeta_A}) should not exceed the 
total curvature perturbations of $\zeta \sim
10^{-5}$ on CMB scales, and further since $\epsilon$ is much smaller than unity
during inflation, 
the condition~(\ref{energy3.6}) is tightened on large scales by at least
5~orders of magnitude.
This strengthens the bound~(\ref{B0bound1}) on the inflation scale to 
\begin{equation}
 H_{\mathrm{inf}} \lesssim 10^{-19}\, \mathrm{GeV},
\label{EQ-19}
\end{equation}
for producing $10^{-15}\, \mathrm{G}$ magnetic fields on Mpc scales. 
With an inflationary scale of $H_{\inf} = 10^{-19}\, \mathrm{GeV}$,
even with the help of instantaneous reheating,
the reheat temperature is restricted to
\begin{equation}
 T_{\mathrm{reh}} \lesssim 10^2 \, \mathrm{MeV}. 
\label{50MeV}
\end{equation}
Since $T_{\mathrm{reh}} $ needs to be higher than about
$5 \,\mathrm{MeV}$ in order not to ruin Big Bang
Nucleosynthesis~\cite{Kawasaki:2000en,Hannestad:2004px},
one sees that our upper bound on~$H_{\mathrm{inf}}$ leaves a rather
narrow window for inflationary magnetogenesis.

We further note that even if the energy density of the vector field
is much smaller than~$\rho_{\mathrm{inf}}$, the produced electric fields
can still prevent the growth of magnetic fields by triggering
Schwinger pair creation of charged particles and thus inducing large
conductivity to the universe~\cite{Kobayashi:2014zza}. 
Extremely strong constraints have been obtained for a class of vector
field theories
with the Lagrangian $I(\tau)^2 F_{\mu \nu} F^{\mu \nu }$;
it would be interesting to systematically study the Schwinger
effect constraints on general inflationary magnetogenesis scenarios
using the analyses presented in this section.

\section{Constraints on Quantum Mechanical Production}
\label{sec:lqf}

In the previous section we derived a bound on magnetic fields 
from a time-evolving classical vector field, cf.~(\ref{growing_assump}). 
However the bound can be evaded if large vector fluctuations are
produced at the quantum level, as then the vector does not necessarily
need to grow after becoming classical.
Such large quantum fluctuations can arise, e.g., 
in theories with modified dispersion relations.

In this section, we derive constraints on quantum mechanical
production of the magnetic fields.
After the modes become classical, we will assume that 
the vector fluctuations do {\it not}
grow significantly in time, 
i.e., the fluctuations in the present universe 
are at most comparable to the initial classical amplitude,
\begin{equation}
 \abs{\tilde{A}_k (\tau_0)}  \lesssim
  \abs{\tilde{A}_k (\tau_\star)}.
\end{equation}
Here, $\tau_\star$ represents the time during
inflation when the mode~$k$ becomes classical,
i.e., when quantum uncertainties become negligible. 
As in Section~\ref{sec:time-evolution}, 
we ignore the component
subscripts of the vector and give order-of-magnitude estimates.
Then the present magnetic field amplitude
obeys
\begin{equation}
\PB(\tau_0, k) \lesssim \frac{k^5}{2 \pi^2} \frac{ \abs{\tilde{A}_{k}(\tau_\star)}^2}{a_0^4}.
  \label{eq19}
\end{equation}
In the following we estimate the right hand side by computing the
vector fluctuations
that can be produced quantum mechanically.
The approach we take in Section~\ref{subsec:smooth} is closely 
related to that used in~\cite{Baumann:2011ws} for studying the Lyth bound in the
context of the effective field theory of
inflation~\cite{Creminelli:2006xe,Cheung:2007st} (see also~\cite{Baumann:2011su}.)

\subsection{Smooth Transition from Quantum to Classical}
\label{subsec:smooth}

Let us suppose that in the asymptotic past, the vector field
follows WKB-like solutions with positive frequencies, i.e.
$\tilde{A}_k \propto e^{- i \int \omega_k d \tau } $, up until the time~$\tau_\star$ when the mode becomes classical. 
Here we assume that the WKB solutions smoothly connect to the classical solutions.
Then, during $\tau \leq \tau_\star$, 
a comoving energy~$\omega_k$ of the vector field can be defined as
\begin{equation}
 \abs{\tilde{A}_k'} \sim \omega_k \abs{\tilde{A}_k}.
\end{equation}
We further assume that the dominant time kinetic term of the vector
field during inflation is a two-derivative term,
\begin{equation}
 S = \int d\tau d^3x  \left(
\frac{I^2}{2} A_i' A_i' + \cdots
\right),
\label{timeKin}
\end{equation}
where the dimensionless coefficient~$I$  may or may not vary in
time, but should not be tiny to avoid strong couplings. 
(See discussions below~(\ref{2der-kin}).)
Then the conjugate momentum of the vector field is
\begin{equation}
 \Pi_i = \frac{\partial \mathcal{L}}{\partial A_i'} = I^2 A_i',
\end{equation}
and thus the commutation relation
\begin{equation}
 \left[
A_i (\tau, \boldsymbol{x}), \Pi_j (\tau, \boldsymbol{y})
\right]
= i \delta^{(3)} (\boldsymbol{x} - \boldsymbol{y})
\left( \delta_{ij} + \cdots \right)
\end{equation}
(here $\cdots$ are additional terms that can show up depending on the
gauge choice), 
sets the normalization of the vector field as
\begin{equation}
  \abs{ \tilde{A}_k \, I^2 \tilde{A}_k' }
\sim  \omega_k \, I^2  \abs{\tilde{A}_k}^2
\sim 1.
  \label{uncertainty}
\end{equation}
Thus, introducing the phase velocity,
\begin{equation}
 c_{\mathrm{p}} (k) \equiv \frac{\omega_k}{k},
\end{equation}
the right hand side of (\ref{eq19}) can be rewritten as 
\begin{equation}
\PB(\tau_0, k)  \lesssim
\frac{1}{2 \pi^2 I_\star^2 c_{\mathrm{p} \star}}
\left(  \frac{k}{a_0}\right)^4 ,
\label{eq4.9}
\end{equation}
where the subscript~$\star$ represents quantities at $\tau = \tau_\star$.
One can see from (\ref{uncertainty}) that, for a fixed wave number~$k$,
the quantum fluctuations scale as $\abs{\tilde{A}_k}^2 \propto
c_{\mathrm{p}}^{-1}$,
i.e., larger fluctuations for smaller phase velocity.
However, the phase velocity at the time the mode becomes classical is actually bounded from
below by energy arguments as follows:
For a classical vector field, its kinetic energy sourced by the time kinetic
term~(\ref{timeKin}) cannot exceed the total energy density of the
inflationary universe
(supposing that the contribution~$\rho_{\mathrm{kin}}$ to
the vector field energy density is not cancelled by other terms
in the action), and thus we require
\begin{equation}
 \int \frac{dk}{k}  \frac{k^3}{2 \pi^2} \frac{I^2}{a^4} |\tilde{A}'_{k}|^2 \ll 3 M_p^2 H_{\mathrm{inf}}^2.
\label{eq4.10}
\end{equation}
Hence the contribution to the kinetic energy
from a classical mode 
should also be smaller than the inflationary energy (see also
Footnote~\ref{foot4}),
\begin{equation}
 \left(
\frac{k^3}{2 \pi^2} \frac{I^2}{a^4} \omega_k^2 |\tilde{A}_{k}|^2
\right)_\star
\ll 3 M_p^2
  H_{\mathrm{inf}}^2 \ ,
\end{equation}
where we used $\Delta k \simeq k$.  This, together with the normalization~(\ref{uncertainty})
at~$\tau_\star$, yields a lower bound on the phase velocity,
\begin{equation}
 c_{\mathrm{p}\star} \gtrsim \frac{1}{(6 \pi^2 M_p^2 H_{\mathrm{inf}}^2)^{1/3}} \left( \frac{\omega_k}{a}
					    \right)^{4/3}_{\star} .
\label{cp-bound}
\end{equation}

Combining (\ref{eq4.9}), (\ref{cp-bound}), and also
$I_\star^2 \gtrsim 1$ for avoiding strong couplings, we find
\begin{equation}
 \PB(\tau_0, k)  \lesssim
  \left(\frac{3 M_p^2 H_{\mathrm{inf}}^2}{4 \pi^4} \right)^{1/3 }
  \left( \frac{a}{\omega_{k} } \right)_\star^{4/3  }
 \left( \frac{k}{a_0} \right)^4.
\label{eq4.13}
\end{equation}
One clearly sees that the magnetic bound is stronger
for a higher energy scale~$(\omega_k / a)_\star$. 
Here, since we are interested in magnetogenesis during inflation,
we assume the modes become classical at a scale at least of order the
inflationary scale,
\begin{equation}
 \left( \frac{\omega_k}{a} \right)_\star \gtrsim H_{\mathrm{inf}}.
  \label{choice31}
\end{equation}
Therefore we arrive at
\begin{equation}
\PB(\tau_0, k)  \lesssim 
\left( \frac{3 M_p^2}{4 \pi^4 H_{\mathrm{inf}}^2} \right)^{1/3} 
 \Bigg( \frac{k}{a_0} \Bigg)^4
\sim
\left( 10^{-43}\, \mathrm{G} \right)^2
\left( \frac{10^{-23}\, \mathrm{GeV}}{H_{\mathrm{inf}}} \right)^{2/3}
\left( \frac{k}{a_0}\, \mathrm{Mpc} \right)^4.
\label{bound32}
\end{equation}
The reference value $10^{-23}\, \mathrm{GeV}$ for $H_{\mathrm{inf}}$
gives a reheat temperature of about $5 \,\mathrm{MeV}$
with the help of instantaneous reheating,
and thus is the lowest possible inflation scale compatible
with Big Bang Nucleosynthesis.
Thus we find that the present magnetic field amplitude cannot exceed
$10^{-43}\, \mathrm{G}$ on Mpc scales or larger, without invoking
classical growth of the vector field.
Considerations on the cosmological perturbations can further strengthen
the bound, as is discussed below~(\ref{zeta_A}).

\subsection{Violent Transition from Quantum to Classical}

In the above discussions, the quantum vector fluctuations were assumed to
smoothly connect to classical fluctuations. 
Here, one may wonder whether large magnetic fields can be produced 
if this assumption is relaxed. 
In this subsection, as an example with non-smooth quantum to classical
transitions, we study 
resonant production of the magnetic fields. Then we will give some
general remarks on quantum mechanical magnetogenesis.

\subsubsection{Resonant Production}
\label{subsubsec:res}

Let us assume an oscillatory background with a physical frequency~$f$ (which
may or may not depend on time), and 
consider the following situation:
The vector fluctuations initially obey WKB solutions with positive
comoving frequencies~$\omega_k$,
but $\omega_k/ a$ eventually becomes comparable to~$f$ and the mode
undergoes parametric resonance.
We represent this time by~$\tau_\star$, i.e.,
\begin{equation}
 \left( \frac{\omega_k}{a} \right)_\star = f_\star,
\end{equation}
where the values with the subscript~$\star$ are estimated at~$\tau_\star$.
After passing through the resonance band, i.e. $\tau > \tau_\star$,
the mode is excited and the vector fluctuation~$\tilde{A}_k$ is a sum of
both positive and negative frequency WKB solutions.
At this point the mode has become classical.  The constraints in this section are inspired by constraints on models of resonant particle production discussed in~\cite{Flauger:2010ja, Behbahani:2011it, Flauger:2013hra}.  

In such a scenario, the estimation of the vector
normalization~(\ref{uncertainty}) arising from the uncertainty principle
can break down after the parametric enhancement, 
as the positive and negative frequency solutions can cancel each other in
the commutation relation.
Instead of invoking the uncertainty relation, we can constrain
the vector amplitude by considering the fact that the resonant
production is induced by the oscillating background; thus the energy
density of the produced vector is bounded by the available kinetic
energy of the background. 
Assuming the vector field's dominant time kinetic term during
inflation to be a two-derivative term, 
\begin{equation}
 S = \int d\tau d^3x  \left(
\frac{I^2}{2} A_i' A_i' + \cdots
	\right),
 \label{4.17}
\end{equation}
we bound the contribution to the vector's kinetic energy from a mode~$k$ that
just underwent parametric enhancement as
(we suppose the contribution to the vector energy is not cancelled by
other terms in the action),
\begin{equation}
 \left(
  \frac{k^3}{2 \pi^2}
  \frac{I^2}{a^4}   \abs{\tilde{A}_k'}^2
\right)_\star
< \epsilon_\star  M_p^2
  H_{\mathrm{inf}}^2.
  \label{4.18}
\end{equation}
The right hand side denotes the
background's kinetic energy,\footnote{Here we are imagining the
inflationary universe to be dominated by canonical scalars,
 $ S = \int d^4 x \sqrt{-g} \left\{ 
\frac{1}{2} M_p^2 R
 -\frac{1}{2} (\partial \phi_1)^2
 -  \frac{1}{2} (\partial \phi_2)^2 - \cdots
 - V(\phi_1, \phi_2, \cdots)
 \right\}$,
with one of the scalars oscillating and sourcing the resonant
production of the vector. 
Then it is easy to check that the total kinetic energy of the
homogeneous background is 
$ \sum_{ i}\frac{1}{2} (\phi_i'/a)^2 = \epsilon M_p^2 H^2 $.}
where $\epsilon = -H' / (a H^2)$
is typically much smaller than unity during inflation. 
The left hand side should be considered as an average over some wave
number range around~$k$, as the $\abs{\tilde{A}_k'}$
spectrum produced from parametric resonance can be oscillatory.
(In cases where the spectrum has spiky features instead of oscillations, 
one should carry out the analyses as discussed in
Footnote~\ref{footinapp}.)\footnote{Strictly speaking, the left hand side of
(\ref{4.18}) is the $k$ mode's  
kinetic energy shortly after the resonant production, while the right
hand side is the background's kinetic energy shortly before the resonant
production. However we abuse notation and denote both by the subscript~$\star$.}
The excited vector fluctuation is a linear combination of WKB solutions with
frequencies $\pm \omega_k$,
however we consider its time derivative to satisfy
\begin{equation}
 \abs{\tilde{A}_k'} \sim  \omega_k \abs{\tilde{A}_k}
  \label{4.19}
\end{equation}
in the sense that $\langle \abs{\tilde{A}_k'}^2 \rangle_T \simeq \omega^2_k \langle
\abs{\tilde{A_k}}^2 \rangle_T$ where $\langle \rangle_T$ means averaged
over one period. 
We also note that the background's oscillation frequency should be larger
than the Hubble rate for parametric resonance to happen, 
\begin{equation}
 \left( \frac{\omega_k}{a} \right)_\star = f_\star > H_{\mathrm{inf}}.
  \label{4.20}
\end{equation}
Hence by combining (\ref{4.18}), (\ref{4.19}), (\ref{4.20}),
and further using $I_\star^2 \gtrsim 1$ to avoid strong couplings, we
obtain a bound on the vector fluctuations,
\begin{equation}
  \abs{\tilde{A}_k(\tau_\star)} \lesssim \frac{\sqrt{2 \epsilon_\star}
   \pi M_p \,  a_\star}{ k ^{3/2}} .
\end{equation}
This field bound can be compared to that derived for classical scenarios
in~(\ref{Akbound}), except that now the bound is stronger by 
$\sim \epsilon_\star$ due to the energy bound~(\ref{4.18}).
Therefore, from (\ref{eq19}) we obtain an upper bound on the present
magnetic field strength,
\begin{equation}
\PB(\tau_0, k)  
 \lesssim 
 \epsilon_\star
 M_p^2 \left( \frac{k}{a_0} \right)^2 \left(
	 \frac{a_\star}{a_0} \right)^2
 \leq
 \epsilon_\star
M_p^2 \left( \frac{k}{a_0} \right)^2 
  \left(\frac{ a_{\mathrm{end}}}{a_0} \right)^2,
\end{equation}
which is stronger than the bound (\ref{B0bound1}) for classical
scenarios by a factor of $\epsilon_\star/3$.

\subsubsection{General Remarks}
\label{subsubsec:GR}

The reader will have noticed that the above argument for constraining
resonant production
also applies for the scenarios with smooth quantum to classical transitions
discussed in Subsection~\ref{subsec:smooth}, by replacing
$\epsilon_\star$ with~3.
In fact, as long as the vector fluctuation that has become classical
satisfies
\begin{equation}
 \abs{ \tilde{A}_k'(\tau_\star) } \gtrsim a_\star H_{\mathrm{inf}}
  \abs{ A_k (\tau_\star) },
 \label{4.23}
\end{equation}
then since the classical mode's kinetic energy cannot
exceed the background density~$3 M_p^2 H_{\mathrm{inf}}^2$,
the fluctuation amplitude is constrained and thus the
magnetic bound is obtained as
\begin{equation}
\PB(\tau_0, k)  
 \lesssim 
 3 M_p^2 \left( \frac{k}{a_0} \right)^2 
  \left(\frac{ a_{\mathrm{end}}}{a_0} \right)^2.
\end{equation}
The bound becomes even stronger when considering the cosmological
perturbations, as was discussed below~(\ref{zeta_A}). 

Therefore whether or not the quantum to classical transition is smooth,
unless (\ref{4.23}) is violated,
bounds on quantum mechanically produced magnetic fields are at least as
strong as the bound on classically enhanced magnetic fields.

\section{Conclusions and Outlook}
\label{sec:conc}

In this work, we presented model independent constraints on magnetic field
generation during inflation by focusing on the field bounds
of the vector potential.
We have classified inflationary magnetogenesis models
according to whether large fluctuations of the vector potential are
produced quantum mechanically,
or the vector fluctuations grow in time after becoming classical,
and derived upper bounds on the
resulting magnetic fields for both scenarios.

For classical scenarios we obtained the bound~(\ref{B0bound1}), 
which requires an extremely low scale inflation of
$H_{\mathrm{inf}} \lesssim 10^{-19}\, \mathrm{GeV}$~(\ref{EQ-19})
in order for magnetic fields of $10^{-15}\, \mathrm{G}$ to be
produced on Mpc scales or larger
without spoiling the cosmological perturbations. 
Thus unless reheating and baryogenesis happen within a very narrow
temperature window of
$5 \,  \mathrm{MeV} \lesssim T_{\mathrm{reh}} \lesssim 10^2 \, \mathrm{MeV}$,
classical scenarios cannot produce the intergalactic
magnetic fields suggested by gamma ray observations. 

Quantum mechanical scenarios were found to be even more restricted;
in cases where the quantum vector fluctuations smoothly convert into
classical fluctuations, 
our result (\ref{bound32}) shows that the produced magnetic fields cannot exceed
$10^{-43}\, \mathrm{G}$ on Mpc or larger scales, independently of the
inflation scale.
Even with non-smooth quantum to classical
transitions (such as in resonant production),
as long as the condition~(\ref{4.23}) is satisfied,
the constraints on quantum mechanically produced magnetic fields are at
least as strong as the constraint for classical scenarios.

Our bounds on classically and quantum mechanically produced magnetic fields
apply to generic inflationary magnetogenesis models with a
two-derivative time kinetic term for the vector field. This was
demonstrated explicitly for the particular examples of
the $I^2 FF$ model, 
and also in the model with a variable light speed due to 
spontaneously broken Lorentz invariance. 
Below, we discuss some possible directions along which our bounds may be
ameliorated. 
\begin{itemize}
 \item {\it Time kinetic terms with less/more than two derivatives.} 
       We have restricted our analyses to vector field theories with 
       two-derivative kinetic terms, however if we relax this condition,
       it may be possible to produce larger magnetic fields. The kinetic term
       $I(\tau)^2 A'_i A'_i$ we have studied also incorporates a certain class of
       one-derivative kinetic terms (as can be seen after canonically
       normalizing the field by $A_i^c = I A_i$), but it may be
       interesting to study theories with other forms of one-derivative, or
       higher-derivative kinetic terms.  
       See also \cite{Baumann:2011su} where an example of a one-derivative
       theory is studied. 
 \item {\it Suppression of time kinetic terms.} 
       If the kinetic term $I^2 A'_i A'_i$ is allowed to be strongly
       suppressed during inflation by an extremely tiny
       coefficient~$I^2$, then in principle large
       magnetic fields can be created while keeping the vector kinetic energy
       subdominant, as was done in the original work~\cite{Ratra:1991bn}.
       However, as we mentioned in
       Section~\ref{sec:time-evolution}, there are at least two obvious
       obstacles to be overcome for taking this option: (i)~A tiny~$I^2$
       gives rise to strong couplings with charged particles.
       Strong couplings may be evaded if there are other factors that suppress
       the interactions (such as an additional scalar derivatively
       coupled to vector-fermion interaction
       terms~\cite{Tasinato:2014fia}, though in this case strong
       couplings may appear in other terms~\cite{Ferreira:2014mwa}), or
       if the charged particles are 
       given large mass during inflation (e.g., Higgs
       inflation~\cite{Bezrukov:2007ep} can help in this direction.)
       However we should also note that such (dynamical) fine-tuning of the
       parameters must happen with an incredible precision, 
       as \cite{Kobayashi:2014zza} demonstrated that even if $I^2$ never
       goes below unity, the Schwinger pair production of charged
       particles can easily terminate magnetogenesis. 
       (ii)~Even if one could avoid strong couplings during
       magnetogenesis, it is non-trivial to arrange $I^2$ to go back to
       unity without affecting the produced magnetic fields. 
       (Note that $I^2$ has been defined to become unity by the present
       epoch.) 
 \item {\it Negative contribution to the vector energy density.}
       One may also consider vector terms
       in the action that contribute negatively to the energy density,
       such that the vector's kinetic energy is cancelled out.
       See~\cite{Campanelli:2015jfa} for an attempt along this line,
       and \cite{Fujita:2012rb} for discussions
       on negative interaction energy.
       However it should also be noted that even if the electric fields
       are cancelled out from the energy density, they can still give rise
       to Schwinger production of charged particles and prevent the
       magnetic fields from growing. 
 \item {\it Other possibilities for quantum mechanical production.} Upon
       constraining quantum mechanical scenarios in
       Section~\ref{sec:lqf}, we assumed the quantum fluctuations to
       start in the Bunch--Davies vacuum. However the situation may
       become different if the initial condition is modified, such as in
       an $\alpha$-vacuum~\cite{Mottola:1984ar,Allen:1985ux}, although there is good reason to
        think such theories are not internally consistent~\cite{Banks:2002nv, Assassi:2012et}. One may
       also investigate the possibility of a violent quantum to
       classical transition such that even the condition~(\ref{4.23}) is
       violated. 
 \item {\it Inverse cascade of helical magnetic fields.}
       Parity violating terms in the action produce helical magnetic
       fields, which can transfer power from
       small to large scales in the radiation-dominated
       era~\cite{Anber:2006xt,Durrer:2010mq,Caprini:2014mja}. 
       Such an inverse cascade can relax the magnetic field
       bounds; it would be interesting to systematically analyze the
       range of possibilities with cascading magnetic fields using the
       field range bounds presented in this paper.
 \item {\it Post-inflationary magnetogenesis.}
       Although the inflationary epoch is preferable for creating
       magnetic fields with large correlation lengths, it should also be
       noted that magnetogenesis can happen in other epochs as well.
       In particular, by breaking the conformal invariance of the
       Maxwell theory after inflation, magnetic fields whose wave modes
       were once inside the horizon during inflation can get
       enhanced in the post-inflationary era, up until the time of
       reheating. 
       As was presented in~\cite{Kobayashi:2014sga}, such 
       post-inflationary scenarios of magnetogenesis can produce large 
       magnetic fields on cosmological scales.  
       Along the lines of avoiding the inflationary epoch, one may also
       consider magnetic field generation in different cosmological backgrounds, such
       as in bouncing
       cosmologies~\cite{Creminelli:2006xe,Khoury:2001wf,Brandenberger:2012zb}.  
       We should also mention the possibility of magnetic field production
       during phase transitions~\cite{Vachaspati:1991nm,Cornwall:1997ms}.
\end{itemize}
These ideas also suggest possible extensions of our formalism. In
particular, it should be straightforward to incorporate
magnetic field evolution after inflation, 
or to consider non-inflationary cosmological backgrounds
(e.g. matter-dominated, contracting phase, etc.) during the generation
of magnetic fields.

\section*{Acknowledgments}

We thank Robert Brandenberger, Tomohiro Fujita, Rajeev Kumar Jain,
Shinji Mukohyama, Bharat Ratra, and Pranjal Trivedi for helpful conversations.  D.G. thanks Daniel Baumann, Raphael Flauger and Rafael Porto for collaboration on related topics. D.G. is support by an NSERC discovery grant.  
T.K. thanks the Institut d'Astrophysique de Paris and the organizers
of the conference ``Mini-workshop on gravity and cosmology'' 
for their hospitality while part of this work was carried out.


\appendix

\section{Implications for $I^2FF$ Models}
\label{app:IFF}

We perform concrete calculations of the constraints on the classical
growth of vectors presented in Section~\ref{sec:time-evolution}, 
in an explicit example with a time-dependent coupling to the vector
kinetic term,
\begin{equation}
 S = \int d^4 x \sqrt{-g}
\left( -\frac{I(\tau )^2}{4} F_{\mu \nu } F^{\mu \nu }   \right).
\label{eq:IFF}
\end{equation}
In this model, 
the time dependence of the coupling~$I$ can break the conformal
invariance of the vector field, and 
the magnetic fields can grow when the coupling~$I$
decreases in time. In the original work~\cite{Ratra:1991bn}, $I$ was
considered to be a function of the inflaton field. 
Here we analyze this model along the lines discussed
in~\cite{Fujita:2012rb}, which studied generic bounds on theories of the 
form~(\ref{eq:IFF}). 
However we obtain stronger bounds by optimizing the combination of
multiple constraints; the differences in the calculations will also be
highlighted along the way.

Let us discuss the field range bound of the vector field in
terms of its mode functions~$u_{k}^{(p)}$, where $p = 1,2$ refers
to the two polarization modes.
Since (\ref{eq:IFF}) is a subclass of the theories 
written as (\ref{ABlag}), 
we refer the reader to Appendix~\ref{app:BTD}
for details of the calculations and here we simply list some relations
that are required for the following analyses:
The electromagnetic power spectra (cf. (\ref{BBpower}), (\ref{EEpower})),
\begin{equation}
 \mathcal{P}_B (\tau, k) = \sum_{p = 1,2}  \mathcal{P}_B^{(p)} (\tau, k),
\qquad
 \mathcal{P}_E (\tau, k) = \sum_{p = 1,2}  \mathcal{P}_E^{(p)} (\tau, k),
\end{equation}
are expressed in terms of the mode functions as
\begin{equation}
 \mathcal{P}^{(p)}_B (\tau, k) = \frac{k^5}{2 \pi^2 a(\tau)^4} 
 | u_k^{(p)} (\tau) |^2,
\qquad
 \mathcal{P}_E^{(p)} (\tau, k) = \frac{k^3}{2 \pi^2 a(\tau)^4} 
 | u_k'^{(p)} (\tau) |^2,
\label{eqA.3}
\end{equation}
which also give the energy density of the vector field
(given as the $c = 1$ limit of (\ref{3-2cc})),
\begin{equation}
 \rho_A = 
\sum_{p = 1,2}  
\frac{I^2}{2}
\int \frac{dk}{k} 
\left( \mathcal{P}_B^{(p)} + \mathcal{P}_E^{(p)} \right).
\label{eqA.5}
\end{equation}
The equation of motion of the mode functions is given in~(\ref{maru1});
in particular, the $I'/I$~term, depending on its sign, can lead to a
growth of~$\abs{u_k^{(p)}}$.
On the other hand, when $I$ is constant in time, 
the mode function is a sum of plane waves and thus
$\abs{u_k^{(p)}}$ is basically constant in time.

Let us now consider the time-evolution of the mode functions between 
two moments of time $\tau_i$ and $\tau_f$ during
inflation,
\begin{equation}
 \abs{ u_k^{(p)} (\tau_f) } -  \abs{ u_k^{(p)} (\tau_i) }
 = \int^{\tau_f}_{\tau_i} d\tau  \, 
\frac{d \abs{u_k^{(p)}}}{d \tau }
\leq \int^{\tau_f}_{\tau_i} d\tau  \, 
\biggl| \frac{d u_k^{(p)}}{d \tau } \biggr|
= \int^{\tau_f}_{\tau_i} d\tau  \, 
\left(
\frac{2 \pi^2 a^4}{k^3} \mathcal{P}_E^{(p)}
\right)^{1/2},
\label{eqA.6}
\end{equation}
where we have used the inequality $d \abs{f(x)} / dx \leq \abs{ d f(x) /
dx }$ for an arbitrary complex function~$f(x)$ with a real variable~$x$,
and also (\ref{eqA.3}) upon moving to the far right hand side.\footnote{In~\cite{Fujita:2012rb}, instead of
as~(\ref{eqA.6}), the time-evolution is constrained as follows,
\begin{equation}
\begin{split}
 \abs{ u_k^{(p)} (\tau_f) }^2 -  \abs{ u_k^{(p)} (\tau_i) }^2
 & = 2 \int^{\tau_f}_{\tau_i} d\tau  \, 
\abs{u_k^{(p)}} \frac{d \abs{u_k^{(p)}}}{d \tau }
\\
& \leq 
2 \int^{\tau_f}_{\tau_i} \frac{d\tau}{k}  \, 
 k \abs{u_k^{(p)}}
 \abs{u_k'^{(p)}}
\\
& \leq \int^{\tau_f}_{\tau_i} \frac{d \tau }{k}
\left(  k^2 \abs{u_k^{(p)}}^2 + \abs{u_k'^{(p)}}^2  \right)
= 
 \int^{\tau_f}_{\tau_i} d\tau \, 
 \frac{2 \pi^2 a^4}{k^4}
\left( \mathcal{P}_B^{(p)} + \mathcal{P}_E^{(p)} \right),
\label{140705}
\end{split}
\end{equation}
where $2xy \leq x^2 + y^2$ for real $x,y$ is used in the second inequality.
The integrand in the final expression is proportional to 
the $k$~mode contribution to the energy density~(\ref{eqA.5}),
hence one can proceed using the energy density arguments as
below~(\ref{eqA.8}). 
However, due to the $1/k$ factor introduced in the second line
of~(\ref{140705}), the resulting magnetic field bound 
is weakened on large scales.} 
We take the final time~$\tau_f$ to be at the end of magnetogenesis
such that $\abs{u_k^{(p)}}$ becomes time-independent afterwards.
The initial time~$\tau_i$ is chosen to be
before~$\tau_f$, but after the
$I'/I$ term in the equation of motion~(\ref{maru1}) becomes relevant;
thus we consider the vector fluctuations as classical 
during $\tau_i \leq \tau \leq \tau_f$.\footnote{The notion of
``being classical" can be made more precise by  
computing the $\kappa$~parameter introduced in
Appendix~\ref{subsec:cl-qu}.  
For example, in the case of $I \propto (-\tau)^s$ with $s$
being a real constant,
if $u_k$ starts in the Bunch--Davies vacuum, then $\kappa \simeq 1 $ in
the asymptotic past ($-k \tau \to \infty$) and thus the vector
fluctuations are quantum mechanical. 
In the asymptotic future ($-k \tau \to 0$),
if $\abs{s} > 1/2$, then $\kappa$ decays to zero and the fluctuations
can be treated as classical.}

The vector field's energy density should
be subdominant to that of the inflationary universe, 
i.e.,
\begin{equation}
 \rho_A \ll 3 M_p^2 H_{\mathrm{inf}}^2,
\label{eqA.8}
\end{equation}
and so we further require the electric field contribution to the energy
density, from each classical wave mode/polarization mode (cf. (\ref{eqA.5}))
to be smaller than the background density,\footnote{Instead of
constraining the magnetic power spectrum on certain wave numbers, 
\cite{Fujita:2012rb} focuses on the magnetic power integrated over a
range of~$k$ that is relevant for interpreting gamma ray
observations~\cite{Tavecchio:2010mk,Neronov:1900zz,Ando:2010rb,Taylor:2011bn,Takahashi:2013lba,Chen:2014rsa}. 
Considering such an effective magnetic strength would also be useful
in cases where the spectrum $\mathcal{P}_E^{(p)}$ has spiky features 
that violate the condition~(\ref{eq1118}) in some narrow $k$ ranges.\label{footinapp}}
\begin{equation}
 \frac{I^2}{2} \mathcal{P}_E^{(p)} \ll 3 M_p^2 H_{\mathrm{inf}}^2.
  \label{eq1118}
\end{equation}
This, combined with $I^2 \geq 1$ for avoiding strong couplings, allow
one to transform (\ref{eqA.6}) into
\begin{equation}
 \abs{ u_k^{(p)} (\tau_f) } -  \abs{ u_k^{(p)} (\tau_i) }
< \frac{\sqrt{12} \pi M_p}{k^{3/2}}
\int^{\tau_f}_{\tau_i} d\tau  \, 
a^2 H_{\mathrm{inf}}
= \frac{\sqrt{12} \pi M_p (a_f - a_i)}{k^{3/2}},
\label{eq.A.9}
\end{equation}
where we used the relation $d \tau = da / (a^2 H_{\mathrm{inf}})$
for the conformal time during inflation.

We now focus on cases where the classical vector fluctuations significantly
grow during inflation, thus by choosing $\tau_i$ to be well
before~$ \tau_f$ such that $a_f \gg a_i$, we suppose
\begin{equation}
 \abs{ u_k^{(p)} (\tau_f) } \gg   \abs{ u_k^{(p)} (\tau_i) }
\end{equation}
to be satisfied.
Then the quantities at $\tau_i$ can be ignored in (\ref{eq.A.9}),
giving
\begin{equation}
\abs{ u_k^{(p)} (\tau_f) } 
 < \frac{\sqrt{12} \pi M_p a_f }{k^{3/2}}.
\end{equation}
Since $\mathcal{P}_B$ scales as $\propto a^{-4}$ after the time~$\tau_f$
(cf.~(\ref{eqA.3})), we can obtain an upper bound on the magnetic power
spectrum in the present epoch,
\begin{equation}
 \mathcal{P}_{B} (\tau_0, k) = 
 \mathcal{P}_{B} (\tau_f, k) 
\left( \frac{a_f}{a_0} \right)^4
<
12 M_p^2 \left( \frac{k}{a_0} \right)^2  
\left( \frac{a_f}{a_0} \right)^2
\leq
12 M_p^2 \left( \frac{k}{a_0} \right)^2  
\left( \frac{a_{\mathrm{end}}}{a_0} \right)^2,
\label{eqA.12}
\end{equation}
where we also used $a_f \leq a_{\mathrm{end}}$. 
Considering the post-inflationary universe to be effectively
matter-dominated until reheating as discussed below~(\ref{eq15}), then 
(\ref{eqA.12}) is rewritten as
\begin{equation}
 \mathcal{P}_{B}(\tau_0, k) <
(10^{-15}\, \mathrm{G})^2  
 \left(\frac{k}{a_0}\, \mathrm{Mpc}\right)^2
 \left( \frac{H_{\mathrm{reh}}}{H_{\mathrm{inf}}} \right)^{1/3}
 \frac{6 \times 10^{-14}\, \mathrm{GeV}}{H_{\mathrm{inf}}}.
\end{equation}
Thus we have reproduced our general bound~(\ref{B0bound1}) in the context
of $I^2 FF$ models.

\section{General Speed of Light Models}
\label{app:BTD}

We study vector field theories in backgrounds that break time
diffeomorphisms and give rise to general speed of propagation. 
As in the previous appendix, we will allow the action to depend explicitly on time but will respect spatial diffeomorphisms. 
For this purpose, it will prove convenient to invoke methods used
in~\cite{Creminelli:2006xe,Cheung:2007st} for analyzing the  
effective field theory of inflation.
Later in this appendix, we will also study a simple magnetogenesis model
where the vector field's sound speed (i.e. speed of light) varies in time. 
This will provide an explicit example of the quantum mechanical
production scenarios discussed in Section~\ref{sec:lqf}. 

\subsection{Action in Unitary Gauge}

We consider the time diffeomorphisms of the vector field theory 
to be spontaneously broken by  
a time-dependent background, such as a time-evolving scalar.
Working in the unitary gauge where
the time coordinate~$x^0$ coincides with the preferred time slicing
of the background, the theory can 
include explicit $x^0$-dependences.
Thus we can construct theories of a vector field with broken time
diffeomorphisms, but unbroken gauge invariance, by considering 
Lagrangians that consist of $(F_{\mu \nu}, \, x^0, \, \nabla_\mu)$. 
The indices should be contracted using the metric~$g_{\mu \nu}$ or the
antisymmetric tensor~$\eta_{\mu \nu \sigma \rho}$,
except for that an upper~$0$ index is allowed to be left free, as indices
can always be contracted with $\nabla_\mu x^0 = \delta_\mu^0$.
We are interested in situations where the vector field does not affect
the background universe, and thus the background fluctuations are
neglected.

In this appendix, we only consider the following quadratic terms containing two derivatives, 
\begin{equation}
 S = \int d^4 x \sqrt{-g} 
\left\{
\mathcal{J} (x^0) F_{\mu \nu} F^{\mu \nu} + 
\mathcal{K} (x^0) 
\tensor{F}{^{0}_{\mu}} F^{0 \mu }
	 \right\},
\label{ABlag}
\end{equation}
where $\mathcal{J}(x^0)$ and $\mathcal{K}(x^0)$ are arbitrary functions
of time~$x^0$.
Note that this theory is invariant under the conformal transformation
$g_{\mu \nu } \to \Omega^2 g_{\mu \nu }$ when the functions also
transform as $\mathcal{J} \to \mathcal{J}$ and $\mathcal{K} \to \Omega^2
\mathcal{K}$. 
With a flat FRW background metric~(\ref{FRW}), the action can be rewritten as
\begin{equation}
 S = \int d\tau d^3 x    \, a(\tau)^4 
\left\{
-\frac{I(\tau)^2}{4}
\left(
2 F_{0i} F^{0i} + c(\tau)^2 F_{ij} F^{ij}
\right)
\right\},
\label{varyingc}
\end{equation}
where we have introduced
\begin{equation}
 I(\tau)^2 \equiv 2 \frac{\mathcal{K}(\tau)}{a(\tau)^2} - 4 \mathcal{J}(\tau),
\qquad
 c(\tau)^2 \equiv \left(
1 - \frac{\mathcal{K}(\tau)}{2 a(\tau)^2 \mathcal{J}(\tau)}
	     \right)^{-1}.
 \label{Iandc}
\end{equation}
The expression~(\ref{varyingc}), and also (\ref{S2.7}) below, clearly
show that this is a theory with a variable speed of light, i.e. sound
speed of the vector field. 
Note in particular that the existence of the $\mathcal{K}$ term deviates
the light speed from unity.

Let us decompose the spatial components of the vector field~$A_i$ into irrotational and
incompressible parts,
\begin{equation}
 A_\mu = (A_0, \, \partial_i S + V_i),
\end{equation}
where 
\begin{equation}
 \partial_i V_i = 0.
  \label{iizero}
\end{equation}
Then the action (\ref{varyingc}) can be rewritten as, up to surface terms,
\begin{equation}
 S = \int d\tau d^3 x \, \frac{I(\tau)^2}{2}
\left\{
 V_i' V_i'  - c(\tau)^2 \partial_i V_j \partial_i V_j
+ \left( \partial_i A_0 - \partial_i S'  \right)
  \left( \partial_i A_0 - \partial_i S'  \right)
\right\}. 
\end{equation}
Varying the action in terms of the Lagrange multiplier~$A_0$ gives a
constraint equation, which, by choosing proper boundary conditions, yields
\begin{equation}
 A_0 = S' .
\end{equation}
Thus we arrive at 
\begin{equation}
 S = \int d\tau d^3 x \, \frac{I(\tau)^2}{2}
\left\{
 V_i' V_i'  - c(\tau)^2 \partial_i V_j \partial_i V_j
	 \right\},
\label{S2.7}
\end{equation}
whose equations of motion, i.e. the modified Maxwell's equations, take
the form of 
\begin{equation}
 V_i'' + 2 \frac{I'}{I}V_i' - c^2 \partial^2 V_i = 0,
  \label{ViEoM}
\end{equation}
where $\partial^2 \equiv \partial_i \partial_i$. 
Going to Fourier space, 
\begin{equation}
 V_i (\tau, \boldsymbol{x}) = 
\frac{1}{(2 \pi )^3} \int d^3 k\, 
e^{i \boldsymbol{k \cdot x}} \xi_i (\tau, \boldsymbol{k}),
\label{fourier}
\end{equation}
then $\xi_i k_i  = 0$ should be satisfied from the 
constraint~(\ref{iizero}). 
We express $\xi_i $ as a linear combination of two orthonormal
polarization vectors~$\epsilon_i^{(p)}(\boldsymbol{k})$ with $p = 1,2$,
that satisfy
\begin{equation}
 \epsilon_i^{(p)} (\boldsymbol{k}) \,  k_i = 0, 
\qquad
 \epsilon_i^{(p)} (\boldsymbol{k}) \epsilon_i^{(q)} (\boldsymbol{k})  =
 \delta_{pq}.
\label{2.10}
\end{equation}
Note that from (\ref{2.10}) follows
\begin{equation}
 \sum_{p = 1,2} \epsilon_i^{(p)} (\boldsymbol{k})
\epsilon_j^{(p)} (\boldsymbol{k})
 = \delta_{ij} - \frac{k_i k_j}{k^2},
\label{epepsum}
\end{equation}
where $k \equiv |\boldsymbol{k}|$. 
We remark that, unlike the spacetime indices, we do {\it not} assume implicit summation
over the polarization index~$(p)$.

\subsection{Quantization}

In order to quantize the theory, we promote $V_i$~(\ref{fourier}) to an
operator, 
\begin{equation}
 V_i(\tau, \boldsymbol{x}) = \frac{1}{(2 \pi)^3}
 \sum_{p = 1,2} \int d^3 k \, \epsilon^{(p)}_i (\boldsymbol{k})
\left\{
e^{i \boldsymbol{k \cdot x}}  a_{\boldsymbol{k}}^{(p)} 
u^{(p)}_{\boldsymbol{k}} (\tau) + 
e^{-i \boldsymbol{k \cdot x}} a_{\boldsymbol{k}}^{\dagger (p)}
u^{*(p)}_{\boldsymbol{k}} (\tau)  
\right\},
\label{Viop}
\end{equation}
where $a_{\boldsymbol{k}}^{(p)}$ and $a_{\boldsymbol{k}}^{\dagger (p)}$ 
are respectively annihilation and creation operators satisfying the
commutation relations,
\begin{equation}
 [ a_{\boldsymbol{k}}^{(p)},\,  a_{\boldsymbol{h}}^{(q)} ] =
 [ a_{\boldsymbol{k}}^{\dagger (p)},\,  a_{\boldsymbol{h}}^{\dagger (q)}
 ] = 0,
\qquad 
 [ a_{\boldsymbol{k}}^{(p)},\,  a_{\boldsymbol{h}}^{\dagger (q)} ] = (2
  \pi)^3 \, 
\delta^{pq} \,
\delta^{(3)}  (\boldsymbol{k} - \boldsymbol{h}) .
 \label{eq:commu}
\end{equation}
$u^{(p)}_{\boldsymbol{k}} (\tau) $ is the mode function, in terms of
which the equation of motion~(\ref{ViEoM}) is rewritten as
\begin{equation}
 u''^{(p)}_{\boldsymbol{k}}+ 2\frac{I'}{I} u'^{(p)}_{\boldsymbol{k}}
+ c^2 k^2  u^{(p)}_{\boldsymbol{k}} = 0.
\label{maru1}
\end{equation}
From the action 
$S = \int d\tau d^3x \, \mathcal{L}$ in (\ref{S2.7}),
the conjugate momentum of $V_i$ is 
\begin{equation}
 \Pi_i = \frac{\partial \mathcal{L}}{\partial V_i'} = I^2 V_i',
\end{equation}
for which the commutation relations are imposed as
\begin{equation}
 \left[ V_i(\tau, \boldsymbol{x}),\,  V_j (\tau, \boldsymbol{y}) \right] = 
 \left[ \Pi_i(\tau, \boldsymbol{x}),\,  \Pi_j (\tau, \boldsymbol{y}) \right]
 = 0,
\label{eq:commu2}
\end{equation}
\begin{equation}
 \left[ V_i(\tau, \boldsymbol{x}),\,  \Pi_j (\tau, \boldsymbol{y})
 \right] = i 
\delta^{(3)}   (\boldsymbol{x} - \boldsymbol{y})
\left( \delta_{ij} - \frac{\partial_i \partial_j}{\partial^2} \right).
\label{eq:commu3}
\end{equation}
The relation (\ref{eq:commu3}) can be rewritten 
using (\ref{epepsum}) as
\begin{equation}
 \left[ V_i(\tau, \boldsymbol{x}),\,  \Pi_j (\tau, \boldsymbol{y})
 \right] = 
 \frac{i}{(2\pi)^3}\sum_{p = 1,2}\int d^3 k \, 
 e^{i\boldsymbol{k\cdot}  (\boldsymbol{x - y})}
\epsilon_i^{(p)} (\boldsymbol{k}) \epsilon_j^{(p)} (\boldsymbol{k}).
\label{eq:commu4}
\end{equation}
We choose the polarization vectors such that
\begin{equation}
 \epsilon^{(p)}_i
  (\boldsymbol{k}) = \epsilon^{(p)}_i (-\boldsymbol{k}),
  \label{poleps}
\end{equation}
then one can check
that the commutation relations~(\ref{eq:commu}) are equivalent to 
(\ref{eq:commu2}) and (\ref{eq:commu4})
when the mode function 
is independent of the direction of~$\boldsymbol{k}$, i.e., 
\begin{equation}
 u_{\boldsymbol{k}}^{(p)} = u_k^{(p)},
\label{maru3}
\end{equation}
and obeys the normalization condition
\begin{equation}
 I^2 \left(
u_k^{(p)} u'^{*(p)}_k - u_k^{*(p)} u'^{(p)}_k
\right) = i.
\label{maru2}
\end{equation}

Defining the vacuum state by
$ a_{\boldsymbol{k}}^{(p)} |0 \rangle = 0 $
for $p = 1,2$ and $^{\forall}  \boldsymbol{k}$,
then (\ref{Viop}) and the commutation relation~(\ref{eq:commu})
allow us to compute the correlation functions of the 
electromagnetic fields~(\ref{EBformal}) as
\begin{align}
 \langle B_\mu (\tau, \boldsymbol{x}) B^\mu (\tau, \boldsymbol{y}) \rangle &=
 \int \frac{d^3 k}{4 \pi k^3}  e^{i\boldsymbol{k\cdot}  (\boldsymbol{x
- y})} \mathcal{P}_B (\tau, k),
\label{BBpower}
\\
 \langle E_\mu (\tau, \boldsymbol{x}) E^\mu (\tau, \boldsymbol{y}) \rangle &=
 \int \frac{d^3 k}{4 \pi k^3}  e^{i\boldsymbol{k\cdot}  (\boldsymbol{x
- y})} \mathcal{P}_E (\tau, k),
\label{EEpower}
\end{align}
where the power spectra are expressed in terms of the mode functions as
\begin{align}
 \mathcal{P}_B (\tau, k) &= \frac{k^5}{2 \pi^2 a(\tau)^4} 
\sum_{p=1,2} | u_k^{(p)} (\tau) |^2,
\label{PofB}
\\
 \mathcal{P}_E (\tau, k) &= \frac{k^3}{2 \pi^2 a(\tau)^4} 
\sum_{p=1,2} | u'^{(p)}_k (\tau) |^2.
\label{PofE}
\end{align}

\subsection{Quantum or Classical}
\label{subsec:cl-qu}

The quantum fluctuations of the vector field can transform into
classical fluctuations due to the time-dependent background. 
To see whether the vector fluctuations are behaving as quantum
mechanical or classical, it is useful to compare the size of
the commutator~$\left[ V,\,  \Pi \right]$
with the amplitude~$ \sqrt{ \langle V^2 \rangle \langle \Pi^2 \rangle }$.

For the purpose of discussing the classical or quantum nature of the observable modes, 
we focus on the Fourier components of the 
operator~$V_i$~(\ref{Viop}) and its conjugate
momentum,
\begin{equation}
 \begin{split}
  V_i(\tau, \boldsymbol{x})
  &= \frac{1}{(2 \pi)^3}
  \sum_{p = 1,2} \int d^3 k \,
e^{i \boldsymbol{k \cdot x}}  \epsilon^{(p)}_i (\boldsymbol{k}) \, 
  v^{(p)}_{\boldsymbol{k}} (\tau),
  \\
  \Pi_i(\tau, \boldsymbol{x})
  &= \frac{1}{(2 \pi)^3}
  \sum_{p = 1,2} \int d^3 k \,
e^{i \boldsymbol{k \cdot x}}  \epsilon^{(p)}_i (\boldsymbol{k}) \, 
  \varpi^{(p)}_{\boldsymbol{k}} (\tau).
  \end{split}
\end{equation}
Here one can check that under (\ref{poleps}), (\ref{maru3}),
(\ref{maru2}),
the commutation relations (\ref{eq:commu}),
or (\ref{eq:commu2}) and (\ref{eq:commu4}),
are rewritten in terms of 
\begin{equation}
\begin{split}
  v^{(p)}_{\boldsymbol{k}} (\tau) & =  
a^{(p)}_{\boldsymbol{k}}  u^{(p)}_{\boldsymbol{k}} (\tau) + 
a^{\dagger (p)}_{-\boldsymbol{k}}  u^{*(p)}_{-\boldsymbol{k}} (\tau),
\\
  \varpi^{(p)}_{\boldsymbol{k}} (\tau) & = 
I^2 \left(
a^{(p)}_{\boldsymbol{k}}  u'^{(p)}_{\boldsymbol{k}} (\tau) + 
a^{\dagger (p)}_{-\boldsymbol{k}}  u'^{*(p)}_{-\boldsymbol{k}} (\tau)
\right) ,
\end{split}
\end{equation}
as
\begin{equation}
 [ v^{(p)}_{\boldsymbol{k}} (\tau),\,  v^{(q)}_{\boldsymbol{h}} (\tau) ]
  =
 [ \varpi^{(p)}_{\boldsymbol{k}} (\tau),\,  \varpi^{(q)}_{\boldsymbol{h}} (\tau) ] =
 0,
 \quad \, 
  [ v^{(p)}_{\boldsymbol{k}} (\tau),\,  \varpi^{(q)}_{\boldsymbol{h}} (\tau) ] =
 i (2   \pi)^3 \, 
\delta^{pq} \,
\delta^{(3)}  (\boldsymbol{k} + \boldsymbol{h}) .
\end{equation}
As an indicator of whether the vector fluctuation is quantum or
classical, we introduce the quantity
\begin{equation}
 \kappa 
  \equiv
  \left|
 \frac{
  [ v^{(p)}_{\boldsymbol{k}} ,
 \varpi^{(q)}_{\boldsymbol{h}}  ]^2
 }{
 4 \, 
 \langle
  v^{(p)}_{\boldsymbol{k}} 
  v^{(q)}_{\boldsymbol{h}} 
  \rangle
  \langle
  \varpi^{(p)}_{\boldsymbol{k}} 
  \varpi^{(q)}_{\boldsymbol{h}} 
 \rangle
 }
 \right|
 =
 \frac{1}{
 4 I^4 \, 
 \abs{u_k^{(p)}}^2 \, 
  \abs{u_k'^{(p)}}^2
}.
\label{kappa}
\end{equation}
For quantum vector fluctuations, $\kappa$ is of order unity. 
On the other hand, if $\kappa \ll 1$, the quantum uncertainty can be
neglected and thus the modes can be considered to have `become classical.'

\subsection{Energy-Momentum Tensor}

In order to compute the energy-momentum tensor of the vector field, we
first invoke the St\"uckelberg trick to the unitary gauge
action~(\ref{ABlag}) and restore the time diffeomorphisms
by reintroducing a Nambu-Goldstone (NG) boson;
then we can vary the covariant action in terms of $g_{\mu \nu}$ and
obtain the energy-momentum tensor in the usual way. 

To apply the St\"uckelberg trick, 
we start by carrying out a time coordinate transformation to the action:
\begin{equation}  \label{time-diff}
\begin{split}
& x^0 \to \tilde{x}^0 = x^0 + \xi^0(x),
 \qquad
 x^i \to \tilde{x}^i = x^i,
 \\
& \, \, 
 A_{\mu } (x) \to \tilde{A}_{\mu } (\tilde{x}) =
  A_{\nu } (x) 
  \frac{\partial x^\nu }{\partial \tilde{x}^\mu },
 \qquad
\mathrm{etc.}
\end{split}
\end{equation}
Then, wherever $\xi^0(x)$ explicitly appears in the transformed action,
we replace $\xi^0(x)$ by the NG boson~$\pi(x)$.
For example,
\begin{equation}
\begin{split}
 \tensor{F}{^{0}_{\mu}}(x) F^{0 \mu }(x)
 & \to
  \tensor{\tilde{F}}{^{0}_{\mu}} (\tilde{x})\tilde{F}^{0 \mu }
 (\tilde{x})
 \\
 & =
  \tensor{F}{^{\nu}_{\rho}}(x) 
  \frac{\partial \tilde{x}^0}{\partial x^\nu }
  \frac{\partial x^\rho }{\partial \tilde{x}^\mu }
  F^{\kappa \lambda} (x)
  \frac{\partial \tilde{x}^0}{\partial x^\kappa }
  \frac{\partial \tilde{x}^\mu }{\partial x^\lambda }
 \\
 & =
   \tensor{F}{^{\nu}_{\rho}}(x)
 F^{\kappa \rho } (x) 
\frac{\partial (x^0 + \xi^0(x))}{\partial x^\nu  }
  \frac{\partial (x^0 + \xi^0(x)) }{\partial x^\kappa }
 \\
 &\Rightarrow
   \tensor{F}{^{\nu}_{\rho}}(x)   F^{\kappa \rho} (x)
  \frac{\partial (x^0 + \pi(x)) }{ \partial x^\nu }
 \frac{\partial (x^0 + \pi(x))}{ \partial x^\kappa },
\end{split}
\end{equation}
where in the last line we replaced the $\xi^0$'s by the NG boson~$\pi$.
Carrying out this procedure for the entire action~(\ref{ABlag}) gives
\begin{equation}
 S_\pi = \int d^4 x \sqrt{-g}
  \left\{
   \mathcal{J}(x^0 + \pi) \, F^{\mu \nu} F_{\mu \nu}
   + \mathcal{K}(x^0 + \pi) \, 
   \tensor{F}{^{\mu}_{\rho}} F^{\nu \rho } \, 
   \partial_\mu (x^0 + \pi)
   \partial_\nu (x^0 + \pi) 
	   \right\}.
  \label{Spi}
\end{equation}
Thus we have restored time-diffeomorphisms; 
one can check that this action is invariant under the 
transformation~(\ref{time-diff}), given that the NG boson also
transforms as 
\begin{equation}
 \pi(x) \to \tilde{\pi} (\tilde{x}) = \pi (x) - \xi^0(x). 
\end{equation}
It can also be seen that the unitary gauge action~(\ref{ABlag}) corresponds to 
the gauge choice of $\pi = 0$.

Now that we have obtained the covariant expression~(\ref{Spi}), we can
follow the standard procedure and 
compute the energy-momentum tensor of the vector fields (and the NG
boson) from $S_\pi = \int d^4 x \sqrt{-g} L_\pi$ as
\begin{equation}
 T^{A, \pi}_{\mu \nu} = g_{\mu \nu} L_\pi - 2 \frac{\partial L_\pi }{\partial
  g^{\mu \nu}}. 
\label{momentum-ten}
\end{equation}
After computing~(\ref{momentum-ten}), we go back to the unitary
gauge $\pi = 0$, where the vector field's energy-momentum tensor is
obtained as
\begin{equation}
\begin{split}
 T^{A}_{\mu \nu} &= g_{\mu \nu} \left(
\mathcal{J} F_{\rho \sigma} F^{\rho \sigma}  + \mathcal{K}
 \tensor{F}{^{0}_{\rho}} F^{0 \rho } 
       \right)
 - 4 \mathcal{J} F_{\mu \rho} \tensor{F}{_\nu^\rho}
 \\
 & \quad
 - 2 \mathcal{K}
 \left(
  \tensor{F}{^0_\mu} \tensor{F}{^0_\nu}
  + \tensor{\delta}{_\mu^0} F_{\nu \rho} F^{0 \rho}
 + \tensor{\delta}{_\nu^0} F_{\mu \rho} F^{0 \rho} 
 \right). 
 \end{split}
\end{equation}
Moreover, in a flat FRW background universe~(\ref{FRW}), the energy
density observed by a 
comoving observer with $4$-velocity~$u^\mu$ is
\begin{equation}
\begin{split}
 T^{A}_{\mu \nu} u^\mu u^\nu 
 = -\tensor{T}{^{A}_0^0}
 &=
  - \mathcal{J} F_{\mu \nu } F^{\mu \nu }
 +  \left( 4 \mathcal{J} - \frac{3 \mathcal{K}}{a^2} \right) F_{0 \mu}
 F^{0 \mu }
 \\
& =  \frac{I^2}{2} \left\{
c^2  B_\mu B^\mu
+ (3 - 2 c^2)  E_\mu E^\mu
\right\},
\end{split}
\end{equation}
where upon moving to the second line we have used (\ref{EBformal}) and
(\ref{Iandc}). 
Therefore the vacuum expectation value of the vector field's
energy density can be expressed in terms of the electromagnetic power spectra
(\ref{PofB}) and (\ref{PofE}) as 
\begin{equation}
\rho_{A}  = \langle -\tensor{T}{^A_0^0} (\tau,
 \boldsymbol{x})  \rangle = 
 \frac{I^2}{2} \int \frac{dk}{k}
 \left\{
 c^2 \mathcal{P}_B + (3 - 2 c^2) \mathcal{P}_E 
 \right\}. 
\label{3-2cc}
\end{equation}

\subsection{Case Study: Inflationary Magnetogenesis from Variable Speed
  of Light}
\label{subsec:mag_varing-c}

Let us now study magnetic field generation in the above theory.
The time dependence of the overall coefficient~$I(\tau)$ of the kinetic
terms has the effect of sourcing a friction term 
in the equation of motion~(\ref{maru1}),
and in particular, a decreasing~$I(\tau)$ gives a negative friction
which can enhance the vector field. This model was studied in
Appendix~\ref{app:IFF}.

In this subsection we would like to focus on the effects of the variable
light speed~$c(\tau)$. For this purpose we fix the coefficient of the
kinetic terms to 
\begin{equation}
 I = 1.
\end{equation}
This case will provide an explicit example of the quantum mechanical
scenarios discussed in Section~\ref{subsec:smooth}.

Hereafter we omit the polarization index~$(p)$, as the computations
are identical for either polarizations.
The equation of motion~(\ref{maru1}) of the mode function is now
\begin{equation}
 u_k'' + c^2 k^2 u_k = 0,
  \label{EOMc2}
\end{equation}
and the constraint~(\ref{maru2}) is 
\begin{equation}
 u_k u_k'^* - u_k^* u_k' = i.
  \label{constI1}
\end{equation}
We consider an inflationary background and express the conformal time
during inflation in terms of the constant Hubble rate~$H_{\mathrm{inf}}$ as 
\begin{equation}
 \tau = -\frac{1}{ a H_{\mathrm{inf}}}.
\end{equation}
Moreover, to make the discussions concrete, 
we focus on cases where the light speed scales as a
power-law of the conformal time,
\begin{equation}
 c(\tau) = c_1 \left( \frac{-\tau }{-\tau_1} \right)^n,
\end{equation}
where $c_1$, $\tau_1$, $n$ are real constants and $c_1$ is further
positive.
Then (\ref{EOMc2}) and (\ref{constI1}) are solved for
the mode function as
\begin{equation}
 u_k (\tau)=
  (-\tau)^{1/2}
  \left\{
\alpha H_{\nu}^{(1)}
  \bigl(
  -2 \abs{\nu}\tau k c(\tau)
  \bigr)
  +
 \beta  H_{\nu}^{(2)}
  \bigl(
  -2 \abs{\nu}\tau k c(\tau)
  \bigr)
  \right\},
\end{equation}
where the index~$\nu$ of the Hankel functions is 
\begin{equation}
  \nu = \frac{1}{2 (n+1)},
\end{equation}
and $\alpha$, $\beta$ are constants satisfying
\begin{equation}
 \left| \alpha \right|^2 -  \left| \beta \right|^2 
 = \frac{\pi \nu }{2} .
\end{equation}

In the following we restrict our analysis to the case of
\begin{equation}
 -1 < n < -\frac{1}{2}
\quad (\mathrm{i.e.}\, \, \nu > 1).
\end{equation}
The lower limit $n > -1 $ is from the requirement that 
the modes with fixed wave numbers~$k$ can exit the sound horizon~$ \sim
c/H_{\mathrm{inf}}$. 
On the other hand, the upper limit $n < -1/2$ is required so that the modes
become classical after sound horizon exit, as we will soon see. 

Under this choice of~$n$, 
the mode functions are well approximated by WKB solutions
in the asymptotic past $- \tau k c \to  \infty$,
as is clear from the 
rate of change of the effective frequency~$c k$ in the equation of
motion (\ref{EOMc2}),
\begin{equation}
  \left\{ \frac{(ck)'}{(ck)^2}  \right\}^2 =
    \frac{n^2}{(-\tau k c)^2},
 \qquad
 \frac{(ck)''}{(ck)^3} = \frac{n (n-1)}{(-\tau k c )^2}. 
\end{equation}
To get the positive frequency solution in the asymptotic past, we set
$\beta = 0$.
Then, in the limit of the modes being well inside the sound horizon, 
i.e. $- 2 \nu \tau k c \to \infty$,
the mode function is approximated as 
\begin{equation}
 u_k \simeq \alpha
  (\pi \nu k c)^{-1/2}
  \exp \left\{
  i \left( 
-2 \nu \tau k c
-\frac{\pi }{4} (2 \nu + 1)
 \right)
       \right\}.
\end{equation}
By computing the parameter introduced in~(\ref{kappa}),
one sees that the modes are quantum mechanical,
\begin{equation}
\kappa = 
  \frac{1}{
 4  \, 
 \abs{u_k}^2 \, 
  \abs{u_k'}^2}
\simeq 1.
\end{equation}
The power spectra (\ref{PofB}) and (\ref{PofE}) are obtained as
\begin{equation}
 \mathcal{P}_B \simeq \frac{k^4}{2 \pi^2 a^4 c},
  \qquad
 \mathcal{P}_E \simeq \frac{c k^4}{2 \pi^2 a^4}. 
\end{equation}

On the other hand, when the modes are well outside the horizon, 
i.e. $-2 \nu \tau k c  \to 0$, 
then
\begin{gather}
 u_k  \simeq - \frac{i }{\pi } 
\alpha (-\tau)^{1/2}
    \Gamma (\nu)(-\nu \tau k c)^{-\nu},
\\
  u_k'
= -\alpha (-\tau)^{-1/2} (-\tau k c) H^{(1)}_{\nu - 1} 
\left(
-2 \nu \tau k c 
  \right)
\simeq 
\frac{i }{\pi }
\alpha (-\tau)^{-1/2}
\frac{\Gamma(\nu-1)}{\nu }
(-\nu \tau k c)^{- \nu  + 2},
\end{gather} 
giving
\begin{equation}
 \kappa \simeq \frac{\pi^2}{\Gamma (\nu)^2 \Gamma (\nu - 1)^2}
(-\nu \tau k c)^{4 (\nu -1)}.
\end{equation}
Since this $\kappa$ decays to zero as $-2 \nu \tau k c \to 0$, one sees that the vector
fluctuations become classical outside the horizon.\footnote{It is also
interesting to analyze the case of $n > -1/2$ (i.e. $0 < \nu 
< 1$). Here, the modes do not necessarily become classical after horizon
exit, as it can be checked that $\kappa \simeq \sin^2 (\pi \nu )$ for
$-2 \nu \tau k c \to 0$.} 
The electromagnetic power spectra are 
\begin{equation}
 \mathcal{P}_B \simeq \frac{k^4}{2 \pi^2 a^4 c}
\frac{\Gamma (\nu)^2}{\pi } \left(-\nu \tau k c  \right)^{- 2 \nu  + 1},
\qquad
\mathcal{P}_E \simeq 
\frac{c k^4}{2 \pi^2 a^4}
\frac{\Gamma (\nu - 1)^2}{\pi }
(-\nu \tau k c)^{-2 \nu + 3 }.
\label{EBoutside}
\end{equation}
Here note that $(-\nu \tau k c)^{- 2 \nu + 1} / c$ is a constant, and
so the magnetic spectrum scales as $\propto a^{-4}$.  
Representing quantities when the mode~$k$ exits the sound horizon
by~$\star$, i.e., 
\begin{equation}
 -2 \nu \tau_\star k c_\star = 1,
\end{equation}
the electromagnetic power spectra outside the horizon can be roughly approximated
as,
\begin{equation}
 \mathcal{P}_B \sim \frac{k^4}{2 \pi^2 a^4 c_\star},
  \qquad
 \mathcal{P}_E \sim
\frac{c_\star k^4}{2 \pi^2 a^4}
\left(\frac{a}{a_\star}\right)^{-4 (n + \frac{1}{2})}.
\end{equation}
Here we have ignored time independent factors that are set by~$\nu$.
These expressions clearly show that 
the electromagnetic amplitudes are set
by the light speed at horizon exit, and in particular that a small light
speed can enhance the magnetic fields while suppressing the electric
fields. 
At first glance, such a feature seems to suggest the possibility of
magnetic field generation without having to worry about the electric
backreaction or Schwinger effect. 
However, the magnetic fields are actually severely constrained 
because of the vector field's energy density:
From (\ref{3-2cc}), the contribution to the energy
density from a mode~$k$ at its horizon exit is estimated as
\begin{equation}
 \rho_{A\star}(k) \sim
\frac{1}{2} \left\{ 
 c_\star^2  \mathcal{P}_{B\star}(k)
 + (3 - 2 c_\star^2)  \mathcal{P}_{E\star}(k) 
\right\} 
\sim \frac{H_{\mathrm{inf}}^4}{\pi^2 c_\star^3}.
\end{equation}
Here, upon moving to the far right hand side, we have assumed $c_\star^2 \ll
1$, 
otherwise the fluctuations of the vector potential would not be larger
than the quantum vacuum fluctuations in the standard Maxwell theory. 
$\mathcal{P}_B$ redshifts as $\propto a^{-4}$ outside the
horizon, and supposing the same redshifting to continue after inflation,
the magnetic power in the present universe is
\begin{equation}
 \mathcal{P}_{B}(\tau_0, k) \sim   
\frac{k^4}{2 \pi^2 a_0^4 c_\star}
\sim 
\left( 10^{-43} \, \mathrm{G}\right)^2
\left(\frac{k}{a_0}\, \mathrm{Mpc} \right)^4
\left( \frac{ \rho_{A\star}(k) }{3 M_p^2 H_{\mathrm{inf}}^2} \right)^{1/3}
\left( \frac{10^{-23}\, \mathrm{GeV}}{H_{\mathrm{inf}}} \right)^{2/3}.
\label{eqB.51}
\end{equation}
The third factor in the far right hand side
is the energy density ratio between the vector field
and the inflationary background,
and the reference value $10^{-23}\, \mathrm{GeV}$ in the fourth factor
gives the lowest possible value for~$H_{\mathrm{inf}}$.
Therefore one sees that the magnetic fields sourced by a variable
light speed during inflation 
cannot exceed $10^{-43}\, \mathrm{G}$ on Mpc scales or larger in the
present universe. 
Thus we have reproduced our magnetic field bound~(\ref{bound32}) in this
explicit model.


\end{document}